\documentclass[epj]{svjour}
\usepackage{graphicx}
\usepackage[english,german,finnish]{babel}
\newcommand{\Real}{\mathrm{Re}\,}
\newcommand{\Imag}{\mathrm{Im}\,}
\begin{document}
\title{Forward analysis of $\pi$N scattering with an expansion method}

\author{P.~Mets\"a
}                     

\institute{Department of Physical Sciences, P.O.\@ Box 64, FIN-00014
  University of Helsinki, Finland}
\date{Received: date / Revised version: date}
%
\abstract{ The $\pi$N forward scattering data are analyzed using an
  expansion method, where the invariant amplitudes are represented by
  expansions satisfying the forward dispersion relations.  The
  experimental errors of the data are taken into account through the
  covariance matrix of the coefficients of the expansions in a careful
  error analysis.  From the results, some coefficients, $c_{n0}^\pm$,
  of the subthreshold expansions have been calculated with proper
  error bars.
\PACS{
      {13.75.Gx}{Pion-baryon interactions}
     } 
} 
\maketitle
\section{Introduction}
\label{intro}

Forward dispersion relations are a special case in $\pi$N scattering
analysis, because in forward scattering the optical theorem provides a
direct connection to total cross section data.  An expansion method
provides a tool for guaranteeing that the forward dispersion relations
are satisfied.  Since the last forward analysis with the expansion
method~\cite{kaiser}, there has not been very much experimental
activity in the forward $\pi$N scattering.  In particular, there are
no new total cross section data except the very high energy
measurements of the SELEX collaboration~\cite{selex}.  However, there
is new information at the physical threshold from the very precise
pionic hydrogen experiments~\cite{schroder}, and, at low energy, from
some new integrated cross section
measurements~\cite{parttot2,parttot6} which are filling the gap
between the physical threshold and the first total cross section data
points.  Both of these have a direct impact on the subthreshold
expansion.

The aim of the present article is to construct invariant amplitudes
$C^\pm$ at $t=0$ to constrain a phase shift analysis with fixed-$t$
analyticity.  Furthermore, information on $\Imag C^-(\omega)$ can be
used to study the Goldberger-Miyazawa-Oehme sum rule~\cite{gmo1,gmo2}.
The expansion method is briefly explained in sect.~\ref{sec:pieexp},
the experimental input is described in sect.~\ref{sec:input}, the
minimizations are discussed in sect.~\ref{sec:results} together with
the evaluation of the subthreshold expansion coefficients.  In
sect.~\ref{sec:conclusions} the conclusions are drawn.

\section{The expansion method}
\label{sec:pieexp}

The isospin even pion-nucleon $C$-amplitude satisfies the
forward dispersion relation~\cite{hoehler}
\begin{eqnarray}
  \label{eq:dispersion}
  & &\Real C^+(\nu,t=0) = -\frac{g^2}{m}\frac{\nu^2}{\nu^2-\nu_B^2}+
  \nonumber \\
  & &
      + \frac{2\nu^2}{\pi} P\!\!\!\!\!\!\int_{\nu_T}^\infty 
      \frac{\mathrm{d}\nu'}{\nu'}  
      \frac{\Imag C^+(\nu',t=0)}{\nu^{'2}-\nu^2} \nonumber \\
  & & 
  + C^+(\nu=0,t=0).
\end{eqnarray}
Here $\nu$ is the crossing antisymmetric Mandelstam variable
$\nu=(s-u)/4m=\omega+t/4m$ with $\omega$ denoting the pion total
laboratory energy, $g^2$ is the $\pi$N coupling constant,
$\nu_B=-\mu^2/2m$ and $C^+(\nu=0,t=0)$ is the subtraction constant.
Everywhere we denote the proton mass by $m$ and the charged pion mass
by $\mu$.  In practice, equations like (\ref{eq:dispersion}) are
difficult to use as constraints, because the principal value
integration is difficult to handle with experimental input.  In
particular, integrating over experimental data is unreliable and
propagating the errors of the input to the error bars of the output is
very difficult.  Instead, we express the isospin even and odd
amplitudes using \foreignlanguage{finnish}{Pietarinen}'s versions of
the expansions~\cite{handbook,pie1972} at $t=0$
\begin{equation}
  \label{eq:pieexp}
  C^\pm(\nu) = C^\pm_N(\nu) + 
  H^\pm(Z)\sum_{k=1}^N c^\pm_k [Z(\nu)]^{k-1},
\end{equation}
where the base function $Z(\nu)$ is a conformal mapping
\begin{equation}
  \label{eq:basefunction}
  Z(\nu)=\frac{\alpha-\sqrt{\nu^2_T-\nu^2}}{\alpha+\sqrt{\nu^2_T-\nu^2}},
\end{equation}
which maps the physical cut to the upper half of the unit circle in
such a way, that the threshold is mapped to $(1,0)$ and the infinity
to $(-1,0)$.  The threshold of the cut in the forward direction is
$\nu_T=\mu$ and the parameter $\alpha$ controls which energy is mapped
to $(0,1)$; there the pion laboratory momentum
$p_\mathrm{lab}=\alpha$.  The numerical value of $\alpha$
in~(\ref{eq:basefunction}) is not crucial, but it is fixed to
$\alpha=0.72$~GeV, which seems to give the most rapid
convergence~\cite{pie1972}.  By using $Z(\nu)$ as the base function,
the analyticity structure of the invariant amplitudes is a built-in
feature and not only a constraint, so the resulting amplitudes will
satisfy the dispersion relations exactly.  In the expansion method
there is also the great advantage, that experimental input for both
real and imaginary parts with their errors can be used simultaneously.
The pole terms~\cite{handbook,pie1976}
\begin{eqnarray}
  \label{eq:CNs}
  C^+_N(\nu) &=& 
  -\frac{\mu^2g^2}{2m}\left(\frac{1}{m^2-s}+\frac{1}{m^2-u}\right),
  \nonumber \\
  C^-_N(\nu) &=& 
  \frac{(s-u)g^2}{4m}\left(\frac{1}{m^2-s}+\frac{1}{m^2-u}\right)
\end{eqnarray}
are treated separately and the assumed high
energy behaviour in the forward direction is taken care of by the
functions $H^\pm(Z(\nu))$~\cite{handbook}
\begin{eqnarray}
  \label{eq:hpms}
  H^+(Z) &=& \frac{1+\left\{r_1 \log[r_2(1+Z)]\right\}^2}{1+Z}, \nonumber \\
  H^-(Z) &=& 4 m \nu (1+Z)^{1-\alpha_\rho}, \\
  \mbox{where }  r_1&=&0.144, \ r_2=44.154 \ ,\mbox{ and } \alpha_\rho=0.48.
  \nonumber
\end{eqnarray}
The number of terms in eq.~(\ref{eq:pieexp}), \emph{i.e.} $N$, was
taken to be 40 in the earlier work~\cite{pie1972,pie1976}, but here we
take $N=100$, which should fully guarantee that the truncation error
is negligible.

An essential ingredient of the expansion technique is the convergence
test function (CTF)~\cite{pie1972,pie1976,pie1973}.  It is not
sufficient to fix the coefficients $c_k^\pm$ in the expansion at
fixed-$t$ by fitting to data, but, in addition, to guarantee the
smoothness of the invariant amplitudes an additional term in the
$\chi^2$ sum is needed.  The convergence test function part takes the
form~\cite{pie1972}
\begin{equation}
  \label{eq:ctf}
  \chi^2_\mathrm{ctf} = W^+_\mathrm{ctf} \sum_{k=1}^N (c_k^+)^2 k^3 + 
  W^-_\mathrm{ctf} \sum_{k=1}^N (c_k^-)^2 k^3,
\end{equation}
where the weights are
\begin{equation}
  \label{eq:ctfweight}
  W^\pm_\mathrm{ctf} = \frac{N}{\sum_{k=1}^N k^3\left[(\bar{c}_k^\pm)^2+
    (\Delta\bar{c}_k^\pm)^2\right]}.
\end{equation}
Here $\bar{c}_k^\pm$ and $\Delta\bar{c}_k^\pm$ denote the expansion
coefficients and their corresponding errors in the isospin even and
odd $C$-amplitudes at the $\chi^2$-minimum, \emph{i.e.} at the best
fit to the data and to the contraints.  So, the minimum has to be
roughly known before the final CTF weights can be calculated.  This
leads to an iterative minimization.

\section{The input}
\label{sec:input}

As input for the fit, we used total cross section
data~\cite{pdg,pdgurl}, integrated cross section
measurements~\cite{parttot2,parttot6}, real-to-imaginary
ratios~\cite{pdg,pdgurl}, real parts of the isoscalar
$D$-amplitude~\cite{wiedner,joram,denz}, the $s$-wave scattering
length $a_{\pi^-p}$ from pionic hydrogen
experiments~\cite{schroder,gmo2} and the scattering length
$a_{\pi^+p}$ from discrete phase shift analysis~\cite{gmo2}.  To begin
with, our full forward data base contains 1098 data points in 142 data
sets covering the laboratory momenta from 0.077~GeV/c to 640~GeV/c in
addition to the threshold values.

\begin{figure}
\resizebox{0.5\textwidth}{!}{%
  \includegraphics{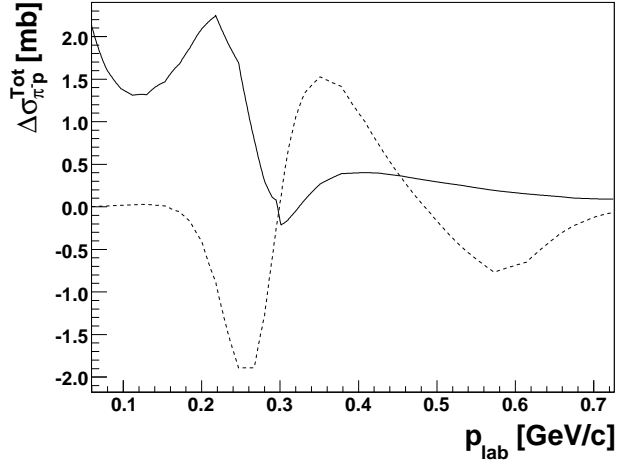}
}
\caption{Electromagnetic (solid line) and $\Delta$-splitting
  corrections (dashed line) for $\sigma^\mathrm{Tot}_{\pi^-p}$.}
\label{fig:emcorpimp}       
\end{figure}

\begin{figure}
\resizebox{0.5\textwidth}{!}{%
  \includegraphics{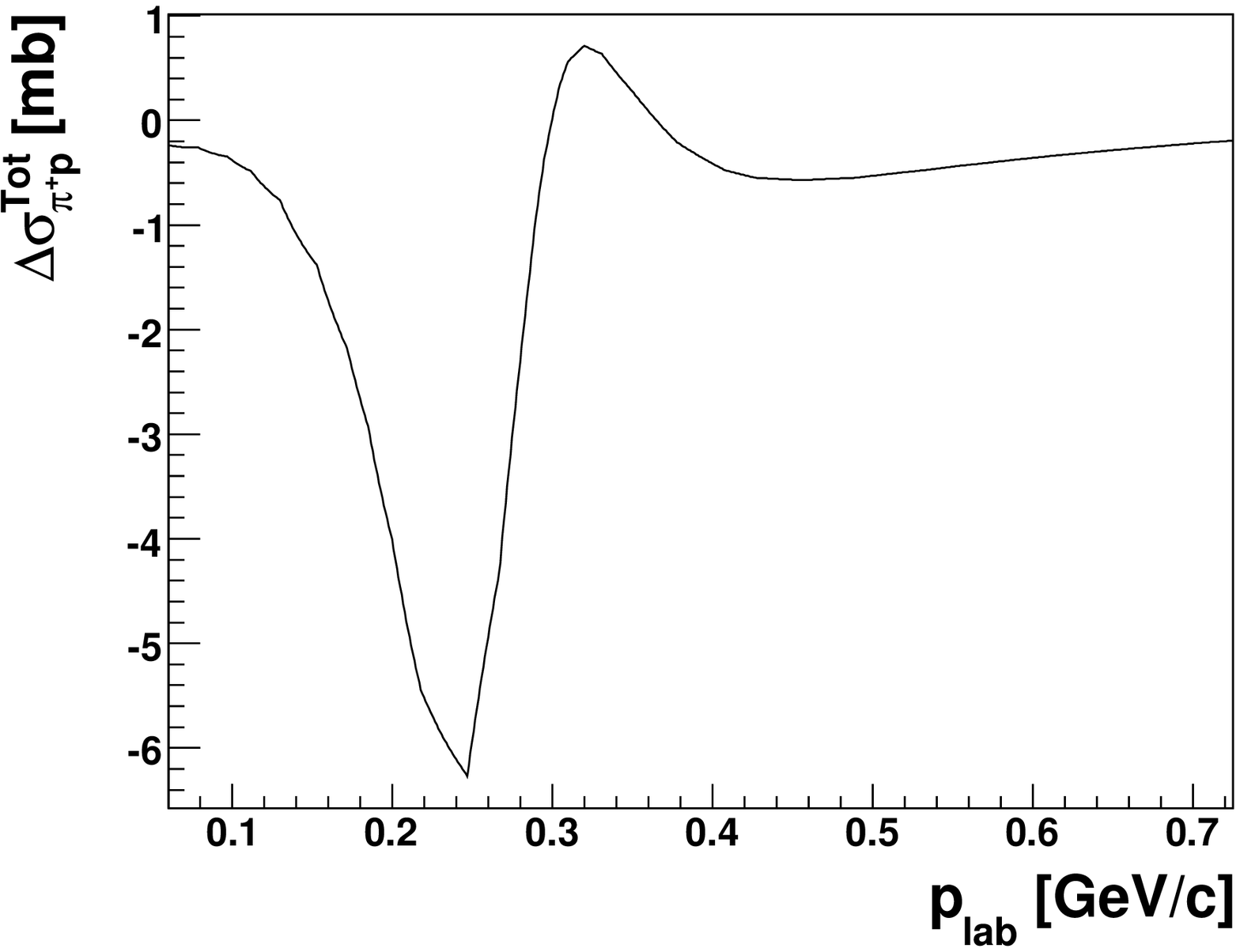}
}
\caption{Electromagnetic correction for
  $\sigma^\mathrm{Tot}_{\pi^+p}$.}
\label{fig:emcorpipp}       
\end{figure}
The electromagnetic effects were removed from the total cross sections
and from the real parts of the $D$-amplitude by the Tromborg
method~\cite{tromborg} for the laboratory momenta
$p_\mathrm{lab}<725$~MeV$\!$/c.  The corrections have been published
only up to 655~MeV/c, but here we employ a smooth extrapolation up to
725~MeV/c.  In order to apply the Tromborg method, an existing partial
wave solution is needed.  We used the KA84 solution~\cite{ka84}, but
the corrections are practically unchanged, if one chooses to take the
KH80~\cite{kh80} or the FA02 solution~\cite{fa02} instead.  The
$\Delta$-splitting was treated by using the $P_{33}$ phase shift
differences from ref.~\cite{abaev}, which are very similar to the
earlier $P_{33}$-corrections of Bugg~\cite{bugg2,bugg3}.  Bugg gives
the corrections for a discrete set of momenta from
$p_\mathrm{lab}=183$~MeV/c to $p_\mathrm{lab}=408$~MeV/c, while with
ref.~\cite{abaev} it is possible to treat all data up to
$p_\mathrm{lab}=725$~MeV/c.  The corrections $\Delta
\sigma^\mathrm{Tot}_{\pi^\pm p}=\sigma^\mathrm{Tot}_{\pi^\pm p} -
\sigma^\mathrm{Tot,hadr}_{\pi^\pm p}$ applied to the measured total
cross sections to obtain the isospin invariant hadronic ones are
displayed in figs.~\ref{fig:emcorpimp} and~\ref{fig:emcorpipp}.  After
applying these corrections, the data were assumed to be purely
hadronic and isospin invariant.  That is reasonable, because any
remaining effects are expected to be considerably smaller than the
experimental errors.  Above $p_\mathrm{lab}=$~725~MeV/c the one photon
exchange picture with the Coulomb phase and form factors was assumed
to be applicable.  For the pionic hydrogen results $\chi$PT-based
electromagnetic corrections~\cite{gasser} were applied to extract the
hadronic quantities.  To make use of the integrated differential cross
section data~\cite{parttot2,parttot6}, they were corrected to hadronic
cross sections integrated over the whole angular range by adding
corrections calculated with the KA84 solution.

The error bars of the total cross sections of Carter
\emph{et~al.}~\cite{carter} were modified by adding the errors due to
the 0.25\% uncertainty in the beam momenta, as explained by
Bugg~\cite{bugg1}.  Also, the corrections adopted by
Giacomelli~\cite{giacomelli} were applied to the total cross sections
of Citron \emph{et~al.}~\cite{citron}.

The forward data alone are not enough to stabilize the low energy
behaviour, \emph{i.e.} the energy range from the threshold up to the
first total cross section data point.  In order to stabilize it
without introducing any bias from earlier solutions, we used our
current partial wave solution to constrain the momentum range 20~MeV/c
$\le p_\mathrm{lab} \le$ 155~MeV/c.  The details of the partial wave
solution will be published elsewhere~\cite{pekkopw}.

\section{Results}
\label{sec:results}

The coefficients $c_k^\pm$ of Pietarinen's
expansions~(\ref{eq:pieexp}) were fixed in a $\chi^2$-minimization
using the program \texttt{MINUIT}~\cite{minuit1,minuit2}.  The actual
minimization was carried out three times, because the convergence test
function method depends on the previously determined minimum.  The
data were allowed to float inside the quoted systematic errors and the
floating factors were searched simultaneously with the coefficients of
Pietarinen's expansions.  In the process of the minimization we had to
discard six data sets, which were too discrepant even after the
renormalization: the $\pi^\pm p$ total cross section sets of Devlin
\emph{et al.}~\cite{devlin65}, two $\pi^+p$ total cross section sets
of Brisson \emph{et~al.}~\cite{brisson59,brisson61}, the $\pi^+p$
total cross section set of Ignatenko \emph{et~al.}~\cite{ignatenko56}
and the $\pi^+p$ total cross section set of Lindenbaum
\emph{et~al.}~\cite{lindenbaum}.  In addition to these, we had to
discard the three lowest data points of Davidson~\emph{et
  al.}~\cite{davidson}.  After excluding these data, we were left with
967 data points in 136 sets and the average $\chi^2$ per data point
was 1.47.
\begin{figure}
\resizebox{0.5\textwidth}{!}{%
  \includegraphics{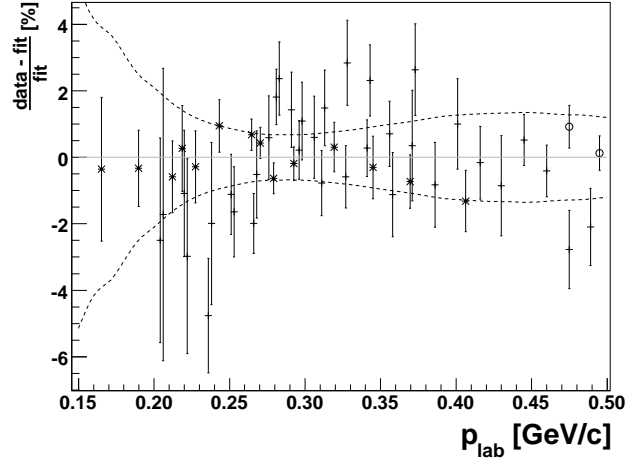}
}
\caption{Difference plot of the total $\pi^-p$ cross section.  The
  dashed lines give the error band of the fit.  The Carter
  data~\protect{\cite{carter}} are marked with crosses, the Pedroni
  data~\protect{\cite{pedroni}} with bars and the Davidson
  data~\protect{\cite{davidson}} with circles.}
\label{fig:tcsm}       
\end{figure}
\begin{figure}
\resizebox{0.5\textwidth}{!}{%
  \includegraphics{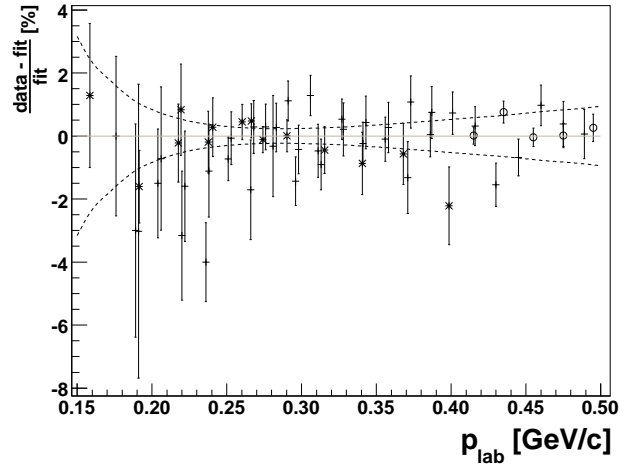}
}
\caption{Difference plot of the total $\pi^+p$ cross section.  For the
  explanation of the symbols for the data, see the caption of
  fig.~\ref{fig:tcsm}.}
\label{fig:tcsp}       
\end{figure}
The differences between the data and the total cross sections
calculated from the fit are plotted in figs.~\ref{fig:tcsm}
and~\ref{fig:tcsp} for the laboratory momentum range
$p_\mathrm{lab}=$150 -- 500~MeV/c.  At higher energy the results are
in good agreement with the earlier work of the Karlsruhe
group~\cite{kaiser}.

Inside the Mandelstam triangle, it is useful to formulate the $\pi$N
amplitudes in terms of the subthreshold expansion with the
pseudovector Born terms subtracted, $\bar{C}^\pm=C^\pm-C^\pm_{N,pv}$.
\begin{table*}
  \caption{The coefficients of the subthreshold expansion of $\bar{C}^+$
    in natural units (powers of $\mu^{-1}$).  In the error estimate the first 
    part is the statistical error, and the second part is the combination of 
    the uncertainty in the coupling constant and the effect of 
    conflicting data sets.  They should be added linearly in order to get 
    the total error.  The Karlsruhe results are 
    from table~2.4.7.1.\@ of ref.~\protect{\cite{hoehler}}
    (N.B. $c^\pm_{n0}=d^\pm_{n0}$).}
\label{tab:dp}       
\begin{tabular}{l|l|l|l}
  \hline
  & \multicolumn{1}{c}{Present analysis} & 
  \multicolumn{1}{c}{Karlsruhe} &
  \multicolumn{1}{c}{SM99~\cite{martin}} \\
  \hline
  $c^+_{00}$ & $-1.20^\dagger\pm0.004\pm0.03$ & $-1.46\pm$0.10  &
             $-1.26\pm$0.02  \\
  $c^+_{10}$ & $\phantom-1.119\pm$0.001$\pm$0.002 &  
              $\phantom-1.12\pm$0.02 &
              $\phantom-1.11\pm$0.02  \\
  $c^+_{20}$ & $\phantom-0.2015\pm$0.0005$\pm$0.0008 & 
              $\phantom-0.200\pm$0.005 &
             $\phantom-0.20\pm$0.01 \\
  $c^+_{30}$ & $\phantom-0.0568\pm$0.0003$\pm$0.0001 & -  &  -  \\
  \hline
\end{tabular}\\
$^\dagger$The value resulting from a calculation with isospin invariance.
\end{table*}
\begin{table*}
  \caption{The coefficients of the subthreshold expansion of $\bar{C}^-/\nu$ 
    in natural units (powers of $\mu^{-1}$).  The error estimates are 
    displayed in the same manner as in table~\ref{tab:dp} 
    except for $c^-_{00}$, where the $g^2$ dependence and 
    the effect of conflicting data are separated.}
\label{tab:dm}       
\begin{tabular}{l|l|l|l}
  \hline\noalign{}
  & \multicolumn{1}{c}{Present analysis} & 
  \multicolumn{1}{c}{Karlsruhe} &
  \multicolumn{1}{c}{SM99~\cite{martin}} \\
  \hline
  $c^-_{00}$ & $\phantom-1.41\pm$0.002$^\mathrm{a}
               \pm0.05^\mathrm{b}\pm0.01^\mathrm{c}$
            & $\phantom-1.53\pm$0.02 & $\phantom-1.43\pm$0.01 \\
  $c^-_{10}$ & $-0.167\pm$0.001$\pm$0.001 & $-0.167\pm$0.005 &
               $-0.16\pm$0.01 \\
  $c^-_{20}$ & $-0.0388\pm$0.0004$\pm$0.0005 & $-0.039\pm$0.002 &
               $-0.04\pm$0.01 \\
  $c^-_{30}$ & $-0.0092\pm$0.0002$\pm$0.0001  & -   & -   \\
  \hline 
\end{tabular}\\
$^\mathrm{a}$The statistical error. \\
$^\mathrm{b}$The $g^2$ dependence, see sect.~\ref{sec:results}. \\
$^\mathrm{c}$Due to conflicting data.
\end{table*}
The subthreshold expansion coefficients are the coefficients in the
expansions based on crossing~\cite{hoehler}
\begin{eqnarray}
  \label{eq:stexp}
  \bar{C}^+(\nu,t) &=& \sum_{n,m} c^+_{nm} \nu^{2n} t^m\nonumber \\
  \bar{C}^-(\nu,t)/\nu &=& \sum_{n,m} c^-_{nm} \nu^{2n} t^m,
\end{eqnarray}
where the isospin odd amplitude is divided by $\nu$ in order to get a
crossing even quantity.  One gets the subthreshold parameters
$c_{n0}^\pm$ by Taylor expanding Pietarinen's representation
(\ref{eq:pieexp}) with the pseudovector Born term subtracted
around $\nu=t=0$
\begin{eqnarray}
  \label{eq:stcofs}
  c^+_{n0} &=& \frac{1}{n!} \sum_{k=1}^N c_k^+ 
  \left.\frac{\partial^n}{\partial(\nu^2)^n} \left[ Z^{k-1}H^+(Z) \right]
    \right|_{\nu=t=0}
  \\
  c^-_{n0} &=& \frac{g^2 \delta_{n0}}{2m^2} + 
  \left.\frac{1}{n!} \sum_{k=1}^N c_k^- 
  \frac{\partial^n}{\partial(\nu^2)^n} \left[ Z^{k-1} H^-(Z)/\nu \right]
  \right|_{\nu=t=0}.
  \nonumber
\end{eqnarray}
The numerical values of the lowest coefficients are given in
tables~\ref{tab:dp} and~\ref{tab:dm} together with the Karlsruhe
values~\cite{hoehler} and the GWU/VPI SM99 values~\cite{martin}.

The statistical errors of the subthreshold expansion
coefficients~(\ref{eq:stcofs}) are calculated by the standard way,
\emph{i.e.}
\begin{equation}
  \label{eq:stater}
  (\Delta c_{n0})^2 = \sum_{k,l} \frac{\partial c_{n0}}{\partial c_k}
  \frac{\partial c_{n0}}{\partial c_l} V_{kl},
\end{equation}
where $V_{kl}$ is the covariance matrix of the Pietarinen
coefficients~$c_k$, calculated by \texttt{MINUIT}.  In Pietarinen's
method, the minimized function is not a pure $\chi^2$-distribution,
but the combination of a $\chi^2$-distribution and the convergence
test function, which generally makes the minimum steeper and gives too
optimistic error bars.  In practice 9.5\% of the probability
distribution at the minimum is due to CTF.  Therefore, in order to get
the proper error bars, we increased the statistical errors of
eq.~(\ref{eq:stater}) by 10\%.

Two sources of systematic errors are studied explicitly.  The effect
of the uncertainty in the coupling constant was estimated by making
calculations with various values\footnote{The pseudoscalar coupling
  constant $g$ is related to the pseudovector coupling $f$ by
  $g^2=4\pi f^2(2m/\mu)^2$.}  in the range
$f^2=0.075\pm0.002$~\cite{gmo2}.  The conflicting data sets are
causing another systematic effect, which was estimated by making the
analysis with different subsets of the data.  The combinations of
these effects are displayed in tables \ref{tab:dp} and \ref{tab:dm}.

The coefficient $c^+_{00}$ can be written as 
\begin{equation}
  \label{eq:c00}
  c^+_{00}=4\pi(1+x)a^+_{0+} + g^2 \frac{x^3}{4-x^2}\frac{1}{\mu}-J^+,
\end{equation}
where $x=\mu/m$, $a^+_{0+}$ is the isoscalar $s$-wave scattering length
and $J^+$ is the integral
\begin{equation}
  \label{eq:Jplus}
  J^+ = \frac{2\mu^2}{\pi} \int_0^\infty \frac{\sigma^+(k)\,
    \mathrm{d}k}{\omega^2}.
\end{equation}
Here $\sigma^+$ is the average $\sigma^+ =
(\sigma^\mathrm{Tot}_{\pi^-p}+\sigma^\mathrm{Tot}_{\pi^+p})/2$.  The
uncertainty in $c^+_{00}$ is mainly due to the first term of
eq.~(\ref{eq:c00}).  The value of the integral $J^+$ is quite stable,
we obtain $J^+ = 1.459\pm0.005$~$\mu^{-1}$, if the integrations are
performed as in ref.~\cite{gmo2}.  The result can be compared with the
\foreignlanguage{german}{Karlsruhe} value~\cite{kochpreprint} $J^+ =
1.478\pm0.010$~$\mu^{-1}$.  If the values for $a_{\pi^\pm p}$ derived
in ref.~\cite{gmo2} are used and isospin invariance is assumed, we
obtain $a^+_{0+}=0.0085\pm0.0016$~$\mu^{-1}$ which gives the value in
table~\ref{tab:dp}.  However, employing the isovector scattering
lengths $a^-_{0+}$ derived from the width measurements of pionic
hydrogen to fix the isoscalar scattering length from
$a^+_{0+}=a_{\pi^-p}-a^-_{0+}$ would give a range of values for
$c^+_{00}$ extending from $-1.17$~$\mu^{-1}$ to $-1.41$~$\mu^{-1}$.
The detailed value depends on which data one wants to fit,
\emph{e.g.}, the Denz \emph{et al.}  $\Real D^+$ data~\cite{denz}
favour a more negative $c^+_{00}$.

To fix $a^+_{0+}$ it is also possible to make use of the measured
level shift in pionic deuterium~\cite{hauser}.  For recent evaluations
of the $\pi^-d$ level shift, see \emph{e.g.}
ref.~\cite{ericson1,beane,doring,meissner}.  Mei\ss ner
\emph{et~al.}~\cite{meissner} include isospin breaking corrections to
$\mathcal{O}(p^2)$ in chiral perturbation theory and give the result
$a^+_{0+}=0.0015\pm0.0022$~$\mu^{-1}$ which would lead to
$c^+_{00}=-1.30$~$\mu^{-1}$, a result well within the above range of
values.

In table~\ref{tab:dm}, the explicit $g^2$ dependence of the parameter
$c^-_{00}$ is given separately as well as the effects due to
conflicting data sets.  For all other parameter values the coupling
constant dependence, the conflicting data effects and the statistical
errors are added linearly.  Beyond $c^\pm_{00}$ all the other
parameters are very stable.

If one uses the value of the coupling constant from Ericson \emph{et
  al.}~\cite{ericson2,ericson3}, $g^2/4\pi=14.07\pm0.17$ instead of
the value of ref.~\cite{gmo2}, the value of the first coefficient of
the isospin odd expansion will be $c^-_{00}=1.49\pm0.03$~$\mu^{-2}$.
The changes in the other coefficients are small.

\section{Conclusions}
\label{sec:conclusions}

The resulting expansions give smooth forward isospin even and odd
$C$-amplitudes, which can be used as a starting point for a phase
shift analysis with fixed-$t$ constraints as well as constraints for a
discrete phase shift analysis~\cite{abaevWIP}.  Also, when the
$\Delta$-splitting corrections have been taken into account, the
hadronic $\pi^\pm p$ total cross sections can be calculated and used
to evaluate the Goldberger-Miyazawa-Oehme sum rule~\cite{gmo2}.

The resulting forward subthreshold expansion co\-ef\-fi\-cients are in
excellent agreement with the earlier evaluations of the Karlsruhe
group except for $c^\pm_{00}$, where the effects due to the coupling
constant and the threshold amplitudes, obtained from pionic hydrogen
experiments, are important.

\begin{acknowledgement}
  I wish to thank E.~Pietarinen for useful exchanges in the early
  phase of the project, A.M.~Green and M.~Sainio for useful comments
  on the manuscript and V.~Abaev for many fruitful discussions.
  Also, I gratefully acknowledge the financial support of the Magnus
  Ehrnrooth Foundation and the Waldemar von Frenckell Foundation.
  This work was supported in part by the EU Contract
  MRTN-CT-2006-035482, FLAVIAnet.
\end{acknowledgement}

%
%

\end{document}